\documentclass[sigconf]{acmart}

\AtBeginDocument{%
  }

\setcopyright{acmlicensed}
\copyrightyear{2026}
\acmYear{2026}
\acmDOI{XXXXXXX.XXXXXXX}
\acmConference[Conference 'XX]{Make sure to enter the correct conference title from your rights confirmation email}{June 03--05, 2018}{Woodstock, NY}
  
\acmISBN{978-1-4503-XXXX-X/2018/06}

\usepackage{algorithm}
\usepackage{algorithmic}
\DeclareMathOperator{\sgn}{sgn}
\usepackage[normalem]{ulem} 

\newcommand{\para}[1]{\noindent \textbf{#1 }}
\newcommand{\sys}{FlowPrefill}

\usepackage{booktabs}   
\usepackage{multirow}   
\usepackage{tabularx}   
\usepackage{adjustbox}
\usepackage[capitalize,noabbrev]{cleveref}

\usepackage[most]{tcolorbox}
\newtcolorbox{insightbox}{
  colback=white,
  colframe=black,
  boxrule=0.6pt,   
  arc=0pt,         
  left=4pt,right=4pt,top=3pt,bottom=3pt,
  boxsep=1pt,
  enhanced,
}
\usepackage[colorinlistoftodos,prependcaption]{todonotes}
\setuptodonotes{inline}

\begin{document}

\title{\sys{}: Decoupling Preemption from Prefill Scheduling Granularity to Mitigate Head-of-Line Blocking in LLM Serving}

\author{Chia-chi Hsieh}
\affiliation{%
  \institution{Tsinghua University}
  \city{Beijing}
  \country{China}}
\email{xiejq24@mails.tsinghua.edu.cn}

\author{Zan Zong}
\authornote{Corresponding authors.}
\affiliation{%
  \institution{University of Science and Technology Beijing}
  \city{Beijing}
  \country{China}}
\email{zongzan@ustb.edu.cn}

\author{Xinyang Chen}
\affiliation{%
  \institution{Tsinghua University}
  \city{Beijing}
  \country{China}}
\email{chenxinyang95@gmail.com}

\author{Jianjiang Li}
\affiliation{%
  \institution{University of Science and Technology Beijing}
  \city{Beijing}
  \country{China}}
\email{lijianjiang@ustb.edu.cn}

\author{Jidong Zhai}
\affiliation{%
  \institution{Tsinghua University}
  \city{Beijing}
  \country{China}}
\email{zhaijidong@tsinghua.edu.cn}

\author{Lijie Wen}
\authornotemark[1]
\affiliation{%
  \institution{Tsinghua University}
  \city{Beijing}
  \country{China}}
\email{wenlj@tsinghua.edu.cn}

\renewcommand{\shortauthors}{Chia-chi Hsieh, Zan Zong, Xinyang Chen, Jianjiang Li, Jidong Zhai, and Lijie Wen}

\begin{abstract}
    The growing demand for large language models (LLMs) requires serving systems to handle many concurrent requests with diverse service level objectives (SLOs). This exacerbates head-of-line (HoL) blocking during the compute-intensive prefill phase, where long-running requests monopolize resources and delay higher-priority ones, leading to widespread time-to-first-token (TTFT) SLO violations. While chunked prefill enables interruptibility, it introduces an inherent trade-off between responsiveness and throughput: reducing chunk size improves response latency but degrades computational efficiency, whereas increasing chunk size maximizes throughput but exacerbates blocking. This necessitates an adaptive preemption mechanism. However, dynamically balancing execution granularity against scheduling overheads remains a key challenge.
    
    In this paper, we propose \sys{}, a TTFT-goodput-optimized serving system that resolves this conflict by decoupling preemption granularity from scheduling frequency. To achieve adaptive prefill scheduling, \sys{} introduces two key innovations: 1) Operator-Level Preemption, which leverages operator boundaries to enable fine-grained execution interruption without the efficiency loss associated with fixed small chunking; and 2) Event-Driven Scheduling, which triggers scheduling decisions only upon request arrival or completion events, thereby supporting efficient preemption responsiveness while minimizing control-plane overhead. Evaluation on real-world production traces shows that \sys{} improves maximum goodput by up to 5.6$\times$ compared to state-of-the-art systems while satisfying heterogeneous SLOs.
\end{abstract}





\maketitle

\section{Introduction}\label{intro}
Large language models (LLMs)~\cite{gpt3, gpt4, touvron2023llama, yang2025qwen3, liu2024deepseek} have demonstrated remarkable generative capabilities and are widely utilized in applications, such as chatbots~\cite{chatgpt, gemini, claude, grok} and code assistants~\cite{codewhisperer, copilot, qwen2-coder}.
As the demand for LLMs grows, service providers~\cite{novita, together, hyperbolic} must meet diverse service level objectives (SLOs) across different users and tasks, making \emph{SLO attainment} a key metric in LLM serving.


\begin{figure}[t]
    \centering
    \includegraphics[width=0.49\textwidth]{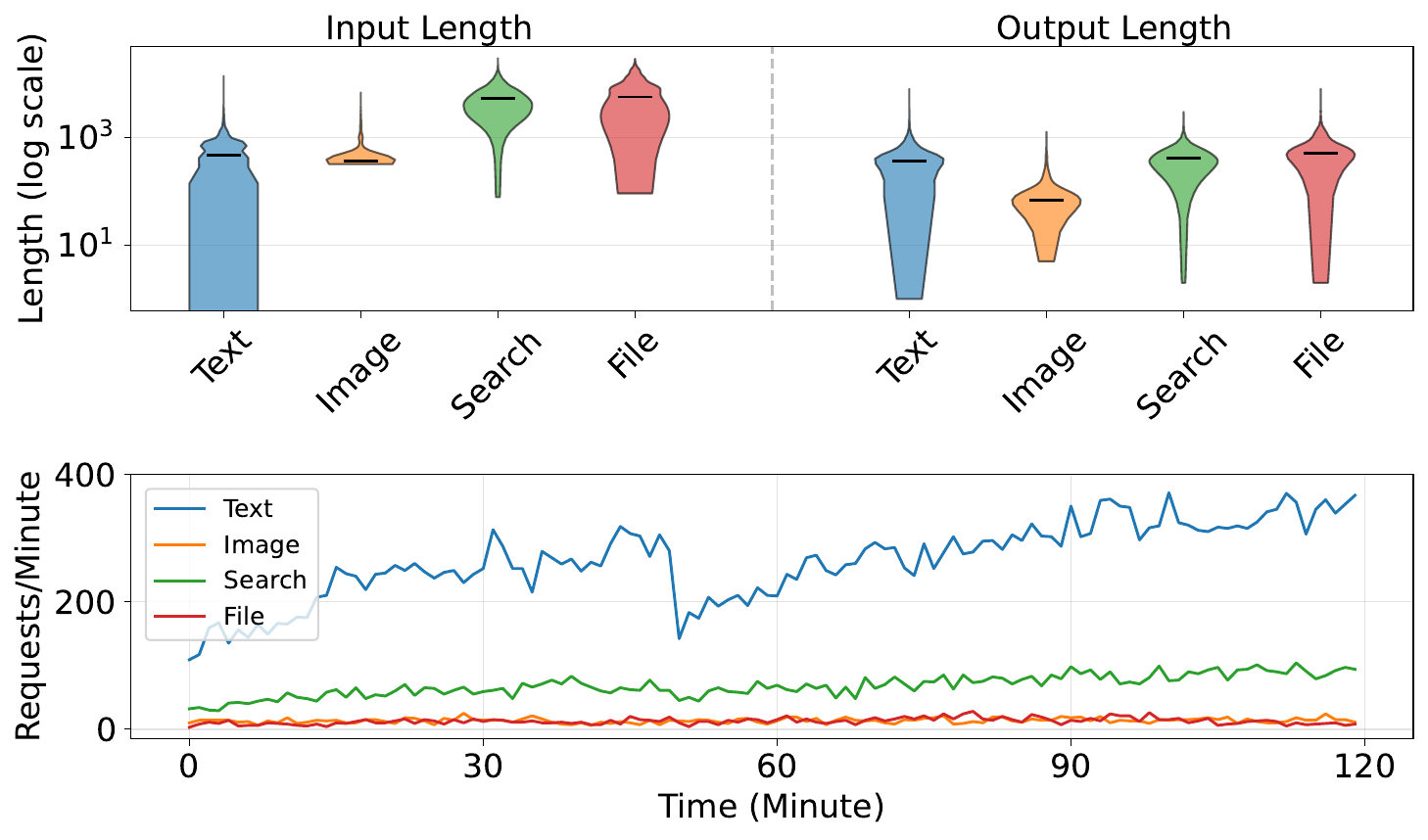}
    \caption{Request length distributions and arrival patterns for different task types in QwenTrace~\cite{qwen-trace}. We explore the possibility of scheduling requests with diverse SLO guarantees.} 
    \label{fig:intro_distribution}
\end{figure}

Time-to-First-Token (TTFT) is a critical metric in LLM serving, as it governs the user's perception of responsiveness and initiates the interactive loop. LLM inference typically consists of two phases with distinct computational profiles: the prefill phase, which processes the input prompt in parallel and is compute-bound; and the decode phase, which generates subsequent tokens autoregressively and is memory-bound. Due to this disparity, existing solutions explore the division of the two phases from both temporal and spatial aspects.
In the temporal division, chunked prefill~\cite{chunkedprefill} splits long prompts into sequential segments to allow decode phases to interleave. However, this approach often fails to guarantee SLOs across different requests.
In the spatial division, prefill-decode (PD) disaggregation~\cite{zhong2024distserve, patel2024splitwise, qin2024mooncake, NVIDIA-Dynamo, liu2025xllm} isolates prefill and decode workloads onto separate hardware to eliminate interference. However, by co-locating the prefill phases of all requests, PD disaggregation can exacerbate resource contention for long prompts.

For service providers, it is essential to support heterogeneous workloads~\cite{azure_trace, qwen-trace, qin2024mooncake}. To maximize service quality, providers must meet diverse latency requirements while maximizing goodput, defined as the maximum sustainable request rate under an SLO attainment goal (e.g., 90\%). For example, a real-world trace illustrated in Figure \ref{fig:intro_distribution} contains multiple task types, each with distinct request length distributions, arrival patterns, and often different SLO requirements~\cite{clockwork, zhong2024distserve, slos-serve, sola, dnnslo}. Task characteristics vary significantly: summarization~\cite{bai2024longbench} often involves long inputs with relaxed SLOs, whereas chatbot interactions~\cite{zheng2023lmsys} typically feature short inputs but demand strict latency. This heterogeneity exacerbates \emph{Head-of-Line (HoL) blocking}: when a long-context prefill monopolizes resources, incoming high-priority requests cannot be scheduled immediately, leading to queuing delays and TTFT SLO violations.

\begin{figure}[t]
    \centering
    \includegraphics[width=0.49\textwidth]{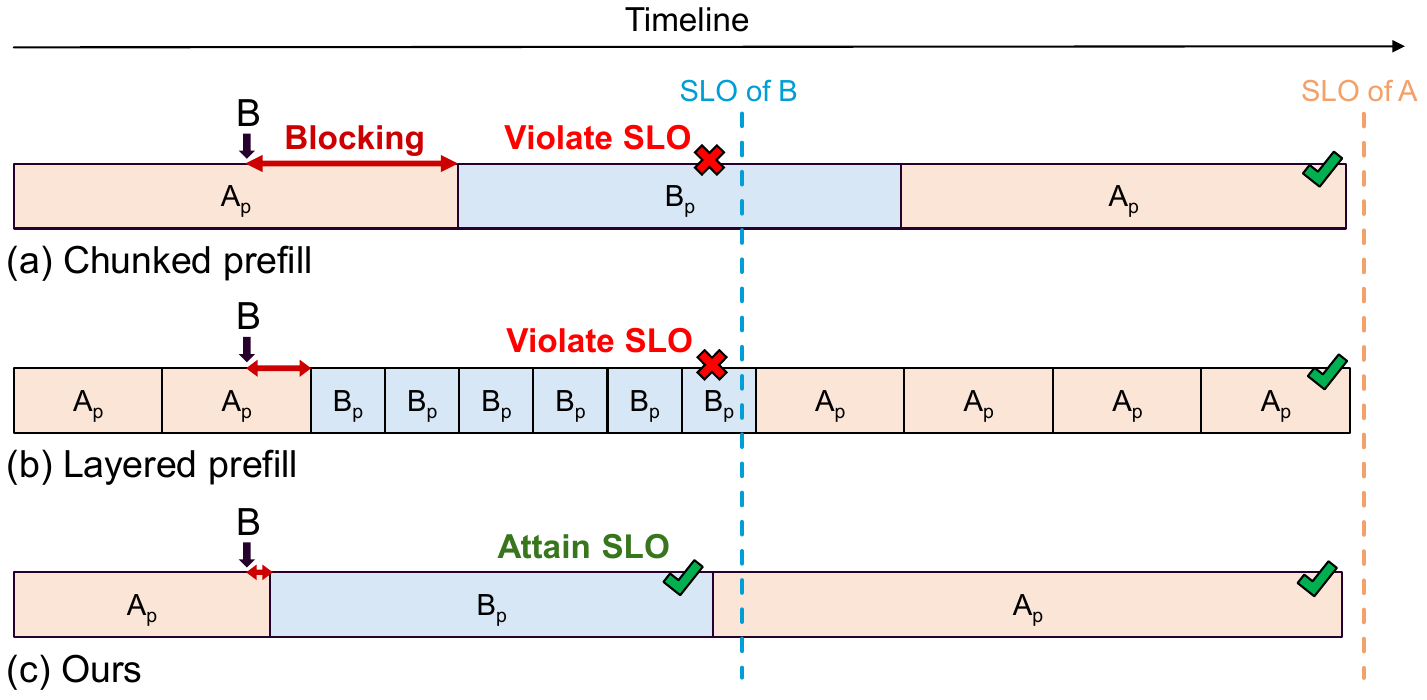}
    \caption{Using different prefill granularities to meet the SLO of time-to-first-token.}
    \label{fig:comparison}
\end{figure}

Existing optimizations such as multi-level queues~\cite{fast-serve, qlm} or output-length prediction~\cite{s3, nips-pred, iclr-pred, qiu-pred}, approximate Shortest-Job-First (SJF) scheduling to alleviate decode-induced HoL blocking. However, these approaches still cannot address HoL blocking caused by long prefill computations.
Recent solutions like chunk-level scheduling~\cite{agrawal2024medha, yu2025prism} introduce preemption at chunk boundaries, often pairing with Earliest-Deadline-First (EDF) policies to improve fairness and SLO adherence (Figure~\ref{fig:comparison}(a)). In addition, layer-level scheduling~\cite{layer-prefill, liao2026laser} offers finer preemption granularity but incurs control-plane overhead proportional to model depth (Figure~\ref{fig:comparison}(b)).
Ultimately, these fixed-size chunk approaches are trapped in a dilemma: \emph{smaller chunks enhance responsiveness but incur higher kernel launch overheads and lower throughput; conversely, larger chunks optimize throughput but exacerbate HoL blocking.}

However, eliminating HoL blocking caused by long-input requests during the prefill phase remains challenging. 
In an ideal scenario, prefill execution would support fine-grained preemption triggered precisely \emph{on-demand}, thereby eliminating redundant scheduling overheads.
To enable efficient LLM serving with strict SLO guarantees, we must address several key challenges:

\para{Balancing preemption granularity and execution efficiency.} Reducing blocking latency requires fine-grained preemption. However, existing approaches rely on fixed-granularity partitioning, leading to a fundamental trade-off: finer granularity increases execution overhead, while coarser granularity exacerbates blocking. An effective solution must therefore support adaptive preemption, enabling interruption at the most appropriate execution boundary without explicitly splitting requests or sacrificing efficiency.

\para{Achieving responsive scheduling without excessive overhead.} Timely response to high-priority requests demands fast scheduling decisions. Yet tightly coupling scheduling with over-segmented execution granularities introduces significant control-plane overhead due to frequent checks. A more desirable approach is to decouple scheduling from execution granularity, enabling timely preemption only when needed while avoiding continuous or unnecessary scheduling overhead.

To address these challenges, we propose \sys{}, a system designed to maximize goodput for online LLM serving. \sys{} is built upon two key ideas: \emph{operator-level preemption} and \emph{event-driven scheduling}, as illustrated in Figure~\ref{fig:comparison}(c). Operator-level preemption enables adaptive runtime preemption at operator boundaries, allowing \sys{} to respond promptly to newly arrived high-priority requests, even during long-running prefill execution. Event-driven scheduling is triggered only on request arrival or completion, enabling SLO-aware prioritization without incurring scheduling overhead proportional to preemption granularity. Consequently, \sys{} is naturally applicable to PD disaggregation systems, where prefill is handled separately, effectively mitigating prefill-induced HoL blocking and maximizing system goodput under heterogeneous SLO requirements.


In summary, our key contributions are:
\begin{itemize}
    \item We reveal a structural limitation where coupling execution granularity with scheduling frequency forces a compromise between throughput and SLO achievement. We propose \sys{}, a goodput-optimized system designed to decouple these two factors for efficient handling of heterogeneous inference requests.
    \item We introduce operator-level preemption, a technique that exploits natural operator boundaries as preemption checks for interruption. This approach minimizes blocking time effectively while avoiding the excessive overhead of small fixed chunks.
    \item We propose an event-driven scheduling framework that separates scheduling decisions from execution granularity. By integrating \emph{Slack-aware EDF} (S-EDF) and \emph{SLO-aware batching} algorithms, our scheduler maximizes goodput and proactively mitigates SLO violations.
    \item We evaluate \sys{} using real-world production traces and state-of-the-art systems such as DistServe \cite{zhong2024distserve} and vLLM \cite{vllm}. It achieves up to 5.6$\times$ higher goodput compared to baselines and can satisfy 3.1$\times$ tighter SLOs, validating its effectiveness in high-demand production environments.
\end{itemize}

\section{Background}
This section details the inference procedure of autoregressive large language models and analyzes the computational characteristics that drive system design decisions. We formally describe the two distinct phases of generation and the fundamental resource constraints associated with each.

\subsection{Transformer-based Generative Inference}
Modern LLMs are built upon the Transformer~\cite{transformer} architecture, which consists of stacked layers containing Self-Attention mechanisms and Feed-Forward Networks (FFN). The inference execution is typically divided into two distinct phases: \emph{prefill} and \emph{decode}. 
First, in the prefill phase, the model processes the entire input sequence (or prompt) $X = \{x_1, x_2, \dots, x_n\}$ in parallel to initialize the Key-Value (KV) cache and predict the first subsequent token. Following this, the system transitions to the decoding phase, which operates autoregressively: the predicted token $x_{n+1}$ is appended to the sequence, and the process repeats iteratively token-by-token until a termination condition is met. For a single Transformer layer, the core operation is Multi-Head Attention (MHA). The input hidden states $X$ are projected into Query ($Q$), Key ($K$), and Value ($V$) matrices using learned weight matrices $W_Q, W_K, W_V \in \mathbb{R}^{d \times d}$: 
\begin{equation}\label{equ:qkv}
\begin{split}
    Q = W_Q X, \quad K = W_K X, \quad V = W_V X
\end{split}
\end{equation}
\noindent{the} attention mechanism then computes a weighted sum of the values based on the similarity between queries and keys. Formally, the output is calculated as: 
\begin{equation}\label{equ:attn}
\begin{split}
\text{Attention}(Q, K, V) = \text{softmax}\left(\frac{QK^T}{\sqrt{d_k}}\right)V
\end{split}
\end{equation}
\noindent{where} $d_k$ is the dimension of the key vectors. Following the attention block, the output passes through a position-wise FFN and residual connections to produce the input for the subsequent layer. This computation is repeated across all $L$ layers of the model.


\subsection{HoL Problem in Disaggregated Serving}

LLM inference comprises two phases with contrasting resource demands. The prefill phase processes the entire input in parallel using Matrix-Matrix Multiplications (GEMMs). It is compute-bound due to high arithmetic intensity, as model weights are reused across all input tokens. Conversely, the decode phase generates tokens autoregressively using Matrix-Vector Multiplications (GEMVs). This phase is memory-bound, as weights must be fetched from HBM for each token, limiting performance by bandwidth rather than computation.

To mitigate resource contention between these phases, modern systems adopt PD disaggregation \cite{zhong2024distserve}, which decouples prefill and decode tasks onto separate GPU instances. This architecture effectively eliminates the interference of compute-intensive prefill bursts on the latency-sensitive decode phase.

However, disaggregation shifts contention entirely to the prefill instances. Since prefill is compute-bound, a single long-context request can monopolize a GPU for hundreds of milliseconds, causing severe HoL blocking for subsequent requests. Consequently, meeting diverse TTFT SLOs becomes challenging, as the scheduler struggles to process heavy workloads without stalling high-priority, short-latency requests.

\section{Motivation}\label{motivation}


In this section, we empirically characterize the performance limitations of prefill execution in online LLM serving. We analyze the tension between preemption granularity and execution efficiency (\S\ref{sec:takeaway_scheduling}), and examine how batching impacts throughput and latency differently across heterogeneous input lengths (\S\ref{sec:takeaway_batching}). Together, these results reveal system-level bounds and workload asymmetries that inform the design of our serving system. Experiments in this section are conducted on an NVIDIA A100-SXM4-40GB GPU.

\subsection{Preemption Granularity \emph{v.s.} Efficiency}\label{sec:takeaway_scheduling}
To mitigate the HoL blocking caused by long prefill requests, existing systems employ fixed chunking strategies, such as Chunked Prefill~\cite{chunkedprefill} and Layered Prefill~\cite{layer-prefill, liao2026laser}. However, our analysis reveals that simply reducing execution granularity introduces a fundamental conflict between responsiveness, computational efficiency, and scheduling overhead.

\begin{figure}[t]
    \centering
    \includegraphics[width=0.49\textwidth]{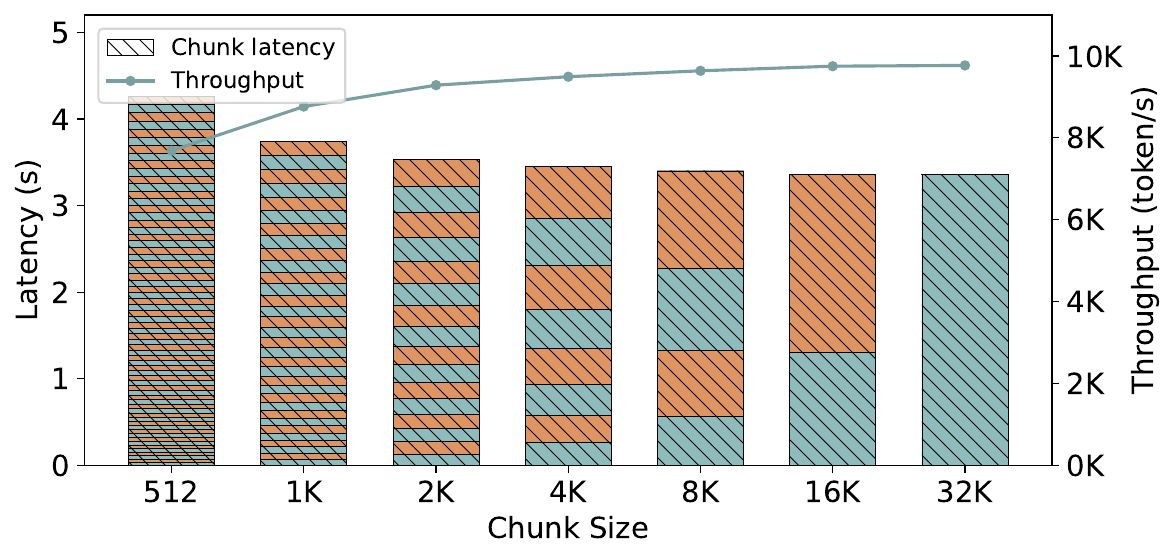}
    \caption{Throughput and latency of chunked prefill for a 32K-token input under different chunk sizes when serving an Llama3-8B. Alternating colors within each bar demarcate individual chunks.}
    \label{fig:chunking}
\end{figure}

\para{Efficiency Degradation in Small Chunking.} Chunked prefill partitions long inputs to enable scheduling at chunk boundaries. As illustrated in Figure~\ref{fig:chunking}, while reducing chunk size improves preemption granularity, it comes at a steep cost to efficiency. Small chunks lead to a significant throughput collapse due to: (1) increased kernel launch overheads; (2) redundant memory accesses to the KV cache (since causal attention~\cite{gpt-1} requires reloading prior keys/values); and (3) under-utilization of device computation. Conversely, large chunks recover throughput but reintroduce substantial blocking latency, failing to satisfy strict TTFT SLOs.

\para{Coupling of Scheduling Decision and Execution Granularity.} Layered prefill avoids the memory overhead of chunking by utilizing the natural boundaries of Transformer layers. However, this approach tightly couples execution granularity with scheduling frequency. If the system performs a scheduling check at every layer to maximize responsiveness, it incurs dramatic control-plane overheads. In practice, high-priority requests do not arrive continuously. Triggering complex scheduling logic at every potential boundary disrupts execution continuity and degrades overall system efficiency—\emph{especially when no preemption is actually required}. Furthermore, for extremely long inputs, even a single layer's execution can take hundreds of milliseconds, implying that layer-level preemption alone imposes an unavoidable granularity bound.

\begin{insightbox}
\para{Takeaway-1:} \textit{Employing fixed, excessively small execution units to minimize blocking degrades hardware efficiency due to redundant memory accesses and kernel launch overheads. Coupling scheduling decisions tightly with these execution boundaries introduces unnecessary control-plane overheads.}
\end{insightbox}




\subsection{Workload Asymmetry in Prefill Batching}\label{sec:takeaway_batching}
Beyond single-request execution, we observe a distinct asymmetry in how batching affects requests of different lengths.

\begin{figure}[t]
  \centering
  \includegraphics[width=\columnwidth]{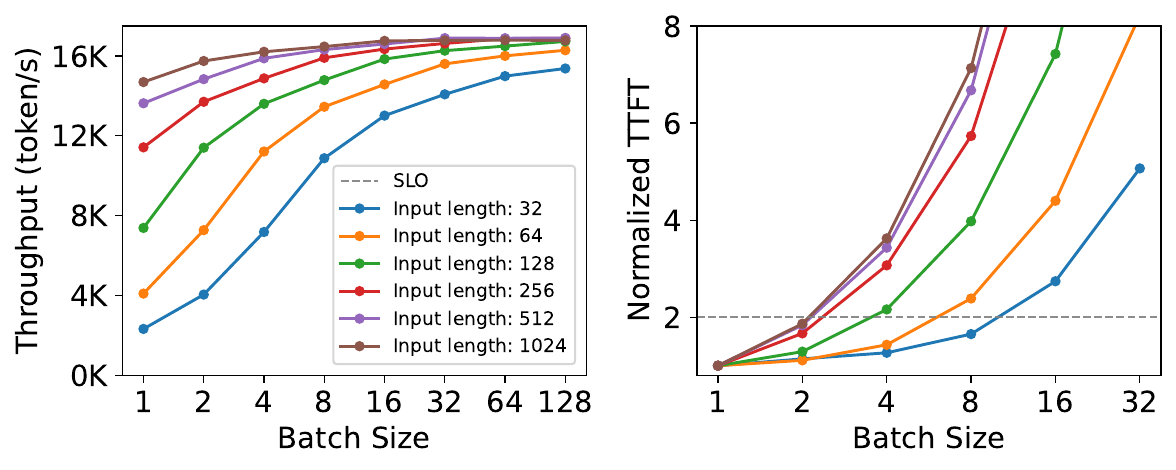}

  \begin{minipage}{0.49\columnwidth}\centering
    \text{(a) Throughput}
  \end{minipage}\hfill
  \begin{minipage}{0.49\columnwidth}\centering
    \text{(b) Normalized TTFT}
  \end{minipage}

  \caption{Throughput and normalized TTFT during the prefill phase for different input lengths when serving an Llama3-8B.}
  \label{fig:batching}
\end{figure}

\para{Throughput Bound for Short Requests.} As shown in Figure~\ref{fig:batching}(a), a single short request (32--256 tokens) cannot fully saturate the GPU's massive parallelism. Consequently, batching multiple short requests is essential. Throughput increases rapidly with batch size before saturating, while latency grows only modestly (Figure~\ref{fig:batching}(b)). This indicates that \emph{for short inputs, aggressive batching significantly boosts hardware utilization and system throughput while maintaining SLO compliance.}

\para{Latency-Bound for Long Requests.} In contrast, long prefill requests are compute- and memory-intensive enough to saturate the GPU individually. For these requests, batching provides minimal throughput gains but causes linear latency inflation due to the aggregated computational load. As a result, long requests in large batches are highly prone to TTFT SLO violations without yielding proportional system-level benefits.

\begin{insightbox}
\para{Takeaway-2:} \textit{It is critical for boosting the throughput of short requests with minimal latency cost, whereas for long requests, it offers negligible throughput gains while significantly increasing the risk of SLO violations.}
\end{insightbox}



\section{Overview}\label{overview}
\begin{figure*}[t]
    \centering
    \includegraphics[width=0.98\textwidth]{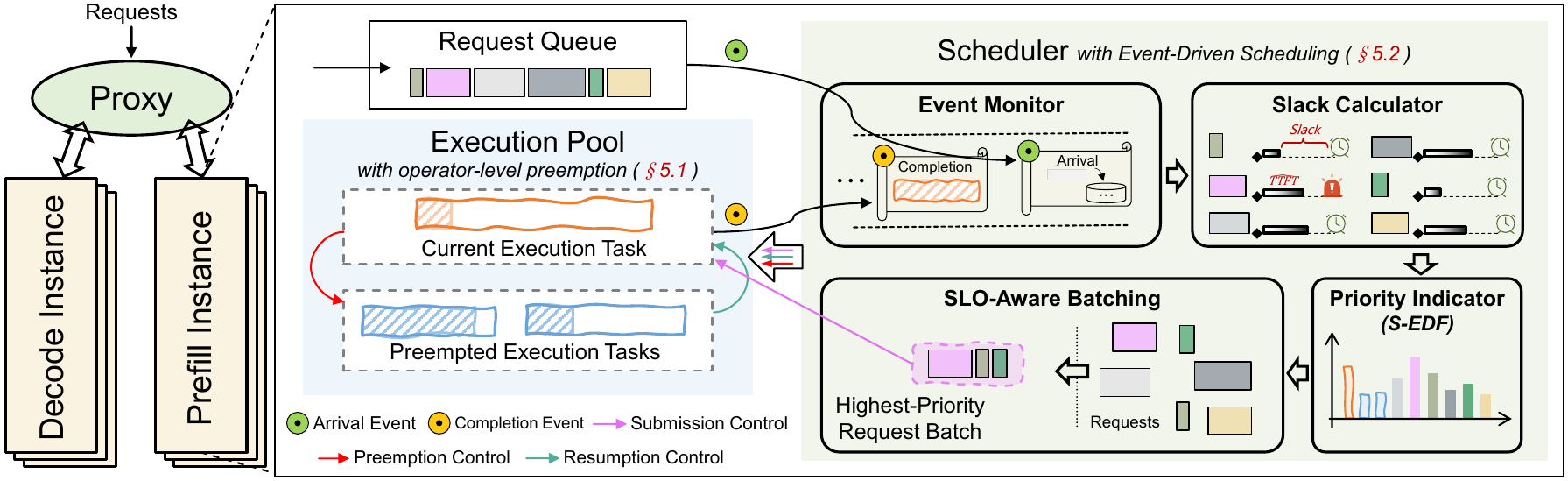}
    \caption{Overview of \sys{}.}
    \label{fig:overview}
\end{figure*}

\sys{} is an LLM serving system optimized for goodput. It directly addresses the trade-off between responsiveness and throughput caused by the fixed setting of granularity, as discussed in \S\ref{motivation}. By decoupling preemption from scheduling granularity, \sys{} enables timely preemption without sacrificing throughput, thereby mitigating severe HoL blocking during the prefill phase and improving online service quality.
As shown in Figure \ref{fig:overview}, a typical serving system consists of three major components: the Proxy, Prefill Instances, and Decode Instances. Our primary optimizations are concentrated in the prefill instances. Each prefill instance comprises three core modules: the Request Queue, the Execution Pool, and the Scheduler. We briefly describe each component below and defer detailed design to \S\ref{Operator-Level Preemption} and \S\ref{Event-driven Scheduling}.

\para{Proxy.} The Proxy acts as the central coordination component of the system. It receives online requests from the frontend, parses them, and sequentially dispatches each request to the prefill and decode instances for inference. Upon inference completion, the Proxy returns the generated outputs to users. When multiple instances are deployed, the Proxy distributes incoming requests using a simple round-robin policy, without considering instance-level load imbalance, which is beyond the scope of this work.

\para{Prefill Instance.} The core optimizations of \sys{} are implemented within the prefill instances to collectively address HoL blocking caused by long-running prefill computation under heterogeneous workloads. To this end, we improve both the scheduling and execution logic of the native inference framework, enabling efficient and safe preemptive scheduling driven by dynamic runtime demands. Each prefill instance consists of the following three components:
\begin{itemize}
\item \textbf{Request Queue.} Each prefill instance maintains a dedicated Request Queue that tracks the logical states of all requests in the system. These states are used to support admission control and scheduling decisions. Incoming requests from the Proxy are enqueued into the Request Queue, while completed requests return their results to the Proxy and are subsequently removed. This design ensures that the Scheduler always operates on an up-to-date and consistent view of all requests.

\item \textbf{Execution Pool.} The Execution Pool manages execution tasks from different request batches. At runtime, it executes at most one task and safely preserves the execution state of preempted tasks until they are resumed. It is responsible only for execution management in response to explicit commands issued by the Scheduler (e.g., \emph{preempt}, \emph{submit}, and \emph{resume}) and does not make request selection or scheduling decisions. The detailed interaction between the Scheduler and the Execution Pool is described in \S\ref{Operator-Level Preemption}.

\item \textbf{Scheduler.} The Scheduler is responsible for request ordering and dispatching. To support dynamic preemption while minimizing scheduling overhead, scheduling is event-driven. An Event Monitor observes an event queue, where actions such as request arrivals and completions enqueue events. The Event Monitor consumes events sequentially, and each event triggers a scheduling round. During each round, the Scheduler consults the request state maintained in the Request Queue and applies the slack calculator, priority indicator, and SLO-aware batching modules to select the highest-priority request batch. Based on this decision, it issues control commands to the Execution Pool, ensuring that the current execution task always corresponds to the highest-priority one. The detailed design is described in \S\ref{Event-driven Scheduling}.
\end{itemize}

\para{Decode Instance.} Decode instances reuse the default execution logic of the native inference framework and schedule decode requests using a first-come-first-served (FCFS) policy. Since decoding optimization is not our primary objective, we concentrate our design and evaluation exclusively on the prefill phase.

\section{Method}
\subsection{Operator-Level Preemption}\label{Operator-Level Preemption}

\begin{figure}[t]
    \centering
    \includegraphics[width=0.46\textwidth]{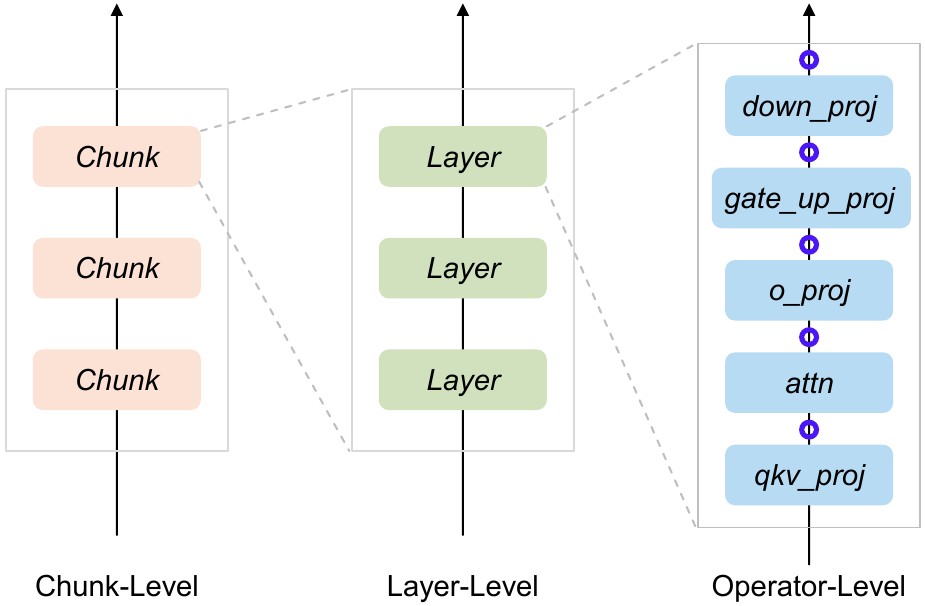}
    \caption{Execution logic is hierarchical, and preemption checks can be placed at different execution boundaries. Preemption checks at the chunk or layer level result in coarser preemption granularity, whereas \sys{} places them at core operator boundaries for the finest-grained preemption. (Blue circles denote preemption checks.)}
    \label{fig:checkpoint}
\end{figure}

Recent approaches attempt to mitigate HoL blocking through chunk- or layer-level scheduling, where preemption is only possible at fixed execution boundaries. However, as discussed in \S\ref{sec:takeaway_scheduling}, this fixed-granularity design introduces a fundamental trade-off: finer granularity improves responsiveness but incurs substantial execution and scheduling overhead, while coarser granularity preserves efficiency at the cost of increased blocking latency.
To overcome this limitation, we propose \emph{operator-level preemption}, which enables preemption immediately after the completion of an operator without request splitting. This design exploits the finest practical execution boundary that preserves kernel semantics, enabling high responsiveness without sacrificing throughput.

To avoid the expensive overhead of process- or context-level preemption~\cite{fan2025gpreempt}, \sys{} adopts cooperative preemption at operator boundaries. During execution, lightweight preemption checks are inserted between operators, allowing the mechanism to generalize across different models and deployment scenarios by flexibly selecting operator boundaries. In this work, which targets dense LLMs, we place preemption checks at the boundaries between core operators, including \texttt{qkv\_proj, attn, o\_proj, gate\_up\_proj, down\_proj}.
As illustrated in Figure~\ref{fig:checkpoint}, preemption checks are performed at every operator boundary to monitor preemption signals from the Scheduler. These operators constitute the smallest practical execution units in LLM architectures, establishing a minimal lower bound on preemption latency. When preemption is requested, the execution runtime adaptively detects the nearest operator boundary and suspends execution upon its completion, enabling timely responses to dynamic preemption demands. Each preemption check consists only of simple concurrency primitive operations, incurring negligible overhead.

\begin{figure}[t]
    \centering
    \includegraphics[width=0.49\textwidth]{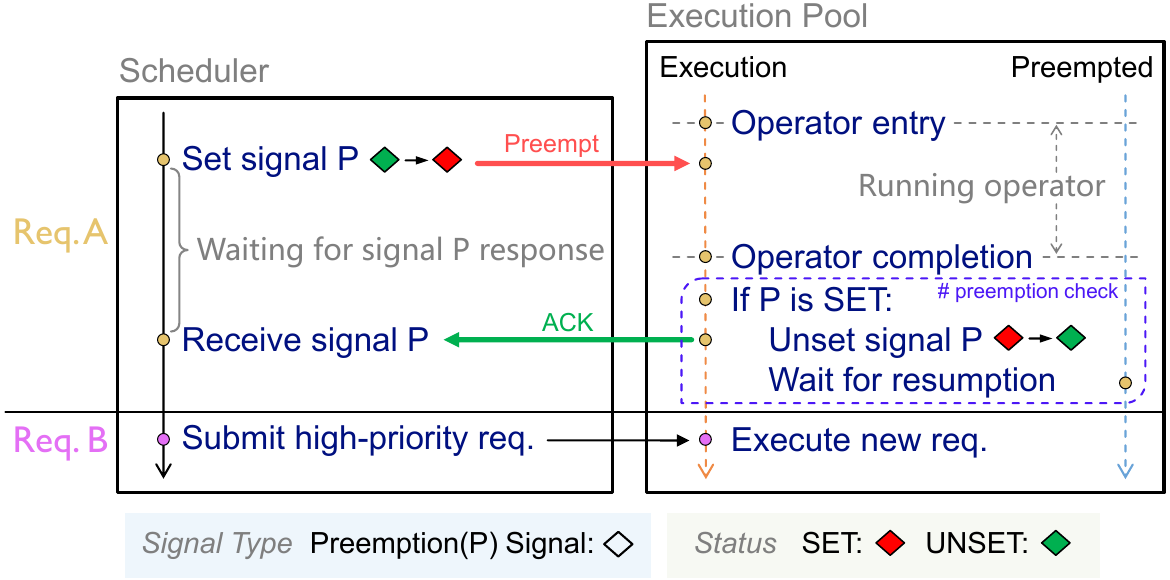}
    \caption{Cooperative preemption process.}
    \label{fig:preemption}
\end{figure}

Once execution is instrumented with preemption checks, it continuously monitors whether the Scheduler has issued a preemption signal. Upon detecting such a signal, execution cooperatively performs preemption and safely suspends the running task. Figure~\ref{fig:preemption} illustrates the cooperative preemption mechanism. When a higher-priority request arrives, the Scheduler sets a preemption signal to notify the Execution Pool and waits for an acknowledgment. Rather than interrupting execution immediately, the runtime performs a preemption check after the current operator completes. If the signal is set, the runtime unsets it to indicate successful preemption and sends an acknowledgment to the Scheduler. It then suspends the current task and moves it to the preempted-task queue, where it awaits resumption by the Scheduler. If the signal is unset, execution proceeds without interruption. Upon receiving the acknowledgment, the Scheduler submits the higher-priority request for execution. By restricting preemption to operator boundaries, \sys{} avoids interrupting in-flight kernels while bounding preemption latency by the execution time of a single operator.

\para{Tensor parallelism supporting.} Serving of large-scale LLMs typically uses tensor parallelism, which runs multiple processes that synchronize via communications. When preemption is required, simply suspending each process is unsafe because faster processes may reach communication points while slower ones are paused, leading to deadlocks. To prevent this, \sys{} uses a synchronized iteration counter among tensor parallel processes and only suspends them when they reach the same counter, ensuring safe preemption without disrupting communication.

\subsection{Event-Driven Scheduling}\label{Event-driven Scheduling}
Operator-level preemption provides the mechanism for safe and efficient interruption. However, supporting heterogeneous workloads with diverse latency requirements further requires a principled scheduling policy that determines \emph{when} to preempt and \emph{which} requests to prioritize. To this end, we propose \emph{event-driven scheduling}, which decouples scheduling decisions from execution granularity while enabling SLO-aware prioritization. Instead of continuously polling for preemption opportunities, scheduling is triggered only by two types of events: request arrival and completion. This design enables timely responses to high-priority requests while avoiding the overhead of frequent scheduling checks.


To reason about request urgency, we propose a slack-aware earliest-deadline-first (S-EDF) scheduling policy. Each request is assigned a priority defined as 
\begin{equation}\label{equ:priority}
\begin{split}
    priority= \frac{\sgn(slack)}{deadline}
\end{split}
\end{equation}
\noindent{where} $slack = deadline - current\_time - \widehat{\text{TTFT}}$, and $\sgn()$ is a signum function returning 1 for positive slack and -1 for negative slack. The $deadline$ is the request arrival time plus its TTFT SLO, $\widehat{\text{TTFT}}$ is the predicted TTFT obtained from a lightweight offline-fitted model.  In other words, the request with the highest priority is the one with the earliest deadline and non-negative slack. By incorporating slack into the priority, S-EDF can proactively deprioritize requests that cannot meet their deadlines, thereby improving the stability of SLO adherence.

\begin{algorithm}[t]
  \caption{SLO-Aware Batching}
  \label{alg:Batching}
  \begin{algorithmic}[1]
    \STATE \textbf{Input:} Highest-priority request $H$, Candidate requests $C$, Batch token budget $G$
    \STATE \textbf{Function} \textsc{SLOawareBatching}($H, C, G$):

    \STATE \quad $B \leftarrow \{H\}$
    \STATE \quad $T_{remain} \leftarrow H.deadline - \text{Timer.time()}$
    \STATE \quad $N \leftarrow H.num\_tokens$
    \STATE \quad \textbf{for} $r \in C$ \textbf{do}
    \STATE \quad \quad $n \leftarrow N + r.num\_tokens$
    
    \STATE \quad \quad $L \leftarrow \text{predict\_latency}(n)$
    \STATE \quad \quad \textbf{if} $T_{remain} > L$ and $n < G$ \textbf{then}
    \STATE \quad \quad \quad $B \leftarrow B \cup \{r\}$
    \STATE \quad \quad \quad $H.num\_tokens \leftarrow n$
    \STATE \quad \quad \quad $N \leftarrow n$

    \STATE \quad  \textbf{return} $H, B$
  \end{algorithmic}
\end{algorithm}

\begin{figure*}[t]
    \centering
    \includegraphics[width=0.99\textwidth]{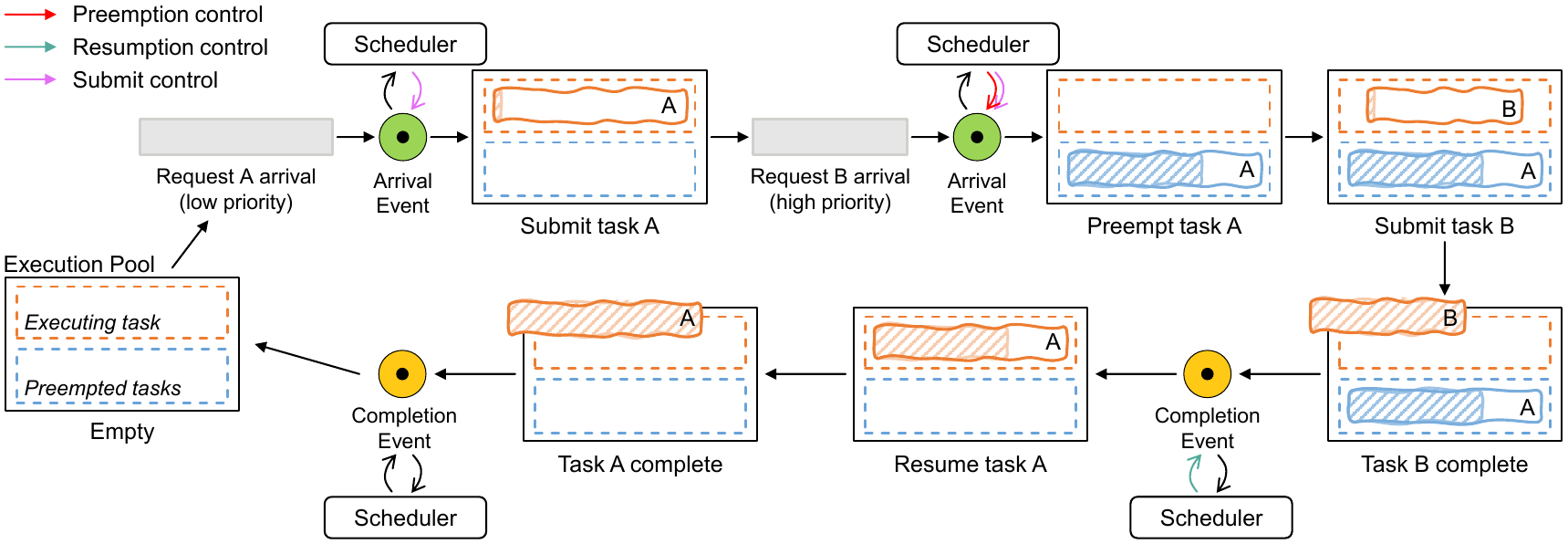}
    \caption{Example of serving two requests with different priorities in \sys{}.}
    \label{fig:workflow}
\end{figure*}
As discussed in \S\ref{sec:takeaway_batching}, existing batching policies do not account for heterogeneous latency requirements. To maximize goodput, we adopt an \emph{SLO-aware batching} strategy (Algorithm~\ref{alg:Batching}), which batches the highest-priority request $H$ with compatible requests only if its remaining time $T_{\text{remain}}$ can accommodate the predicted batch latency $L$ and the batch token budget $G$ is not exceeded. The batch token budget ensures that the benefits of batching short requests are properly captured. When a request is admitted to the batch, the aggregate token count of $H$ is incrementally updated to guide subsequent admission decisions. By allocating a controlled token budget to $H$ within latency constraints, this strategy minimizes the risk of SLO violations while maximizing system throughput.

The complete scheduling procedure is shown in Algorithm~\ref{alg:Scheduling}. At each scheduling step, we first wait for an event (arrival or completion) to occur (line 4). After that, all requests $Q_{all}$ are ranked by descending priority, and the highest-priority request $H$ is selected (lines 5–12). If $H$ is in the waiting queue $Q_w$, indicating that it has not yet been assigned an execution task, we attempt SLO-aware batching (Algorithm~\ref{alg:Batching}) to batch it with compatible requests (lines 13–15). Finally, we check whether the currently running execution $E$ should be preempted, moving it to the preemption queue $Q_p$ if needed, and then either submits a new execution task or resumes a previously preempted task (lines 16–26). By combining event-driven scheduling with operator-level preemption, \sys{} achieves responsive, SLO-aware scheduling while avoiding the overhead of fine-grained, continuously triggered scheduling.

\begin{algorithm}[t]
  \caption{Event-Driven Scheduling}
  \label{alg:Scheduling}
  \begin{algorithmic}[1]
    \STATE \textbf{Input:} Batch token budget $G$
    \STATE Initialize $Q_w \leftarrow \emptyset$,  $Q_p \leftarrow \emptyset $,  $E \leftarrow null$
    
    \STATE \textbf{while} True \textbf{then}
    \STATE \quad Wait for request arrival or completion event
    \STATE \quad $r_w \leftarrow \text{get\_new\_requests()}$
    \STATE \quad $Q_w \leftarrow Q_w \cup r_w$
    \STATE \quad $Q_{all} \leftarrow Q_w \cup Q_p \cup \{E\}$
    \STATE \quad \textbf{if} $Q_{all} = \emptyset$ \textbf{then}
    \STATE \quad \quad \textbf{continue}
    \STATE \quad Sort requests $Q_{all}$ in descending order of $r.\textit{priority}$, yielding 
    \STATE \quad $(r_1, r_2, \ldots)$ with $r_1$ having the highest priority.

    \STATE \quad Initialize $H \leftarrow r_1$, $B \leftarrow \emptyset$
    \STATE \quad \textbf{if} $H \in Q_w$ \textbf{then}
    \STATE \quad \quad $C \leftarrow Q_{all} \setminus Q_p \setminus\{H\}$
    \STATE \quad \quad $H, B \leftarrow$ \hyperref[alg:Batching]{\textsc{SLOawareBatching}}($H, C, G$)

    \STATE \quad \textbf{if} $H \neq E$ \textbf{then}
    \STATE \quad \quad \textbf{if} $E \neq$ null \textbf{then}
    \STATE \quad \quad \quad $\text{preempt\_execution($E$)}$
    \STATE \quad \quad \quad $Q_p \leftarrow Q_p \cup \{E\}$

    \STATE \quad \quad \textbf{if} $B \neq \emptyset$ \textbf{then}
    \STATE \quad \quad \quad $\text{submit\_new\_execution($H, B$)}$
    \STATE \quad \quad \quad $Q_w \leftarrow Q_w \setminus B$
    \STATE \quad \quad \textbf{else}
    \STATE \quad \quad \quad $\text{resume\_execution($H$)}$
    \STATE \quad \quad \quad $Q_p \leftarrow Q_p \setminus \{H\}$

    \STATE \quad \quad $E \leftarrow H$
  \end{algorithmic}
\end{algorithm}

\para{An illustrated example.}
As shown in Figure~\ref{fig:workflow}, we present a simple example to illustrate the event-driven scheduling logic in \sys{}, covering all control commands including \emph{submit}, \emph{preempt}, and \emph{resume}. The example involves two requests, request A with low priority and request B with high priority. Initially, there is no executing task and the set of preempted tasks is empty. When request A arrives, it generates an event that triggers a scheduling round, and since no execution is running, the Scheduler issues a \emph{submit} command to execute request A as task A. When request B arrives, it generates another event and triggers a new scheduling round. Because request B has higher priority, the Scheduler issues a \emph{preempt} command to suspend task A and free the execution slot, followed by a \emph{submit} command to execute request B as task B. After task B completes, its completion generates an event that triggers another scheduling round. Since no higher-priority requests remain, request A becomes the highest-priority request and the Scheduler issues a \emph{resume} command to continue its execution task. Finally, when task A completes, it generates an event that triggers a scheduling round with no runnable requests. Throughout this process, the Scheduler is responsible for issuing control commands, while the Execution Pool carries out the corresponding actions on the execution tasks of the relevant requests, as described in \S\ref{overview}.


\section{Evaluation}
\subsection{Experimental Setup}
\para{Implementation.} We implement \sys{} on vLLM-0.11.2~\cite{vllm}. The Execution Pool uses \texttt{ThreadPoolExecutor} to manage per-request execution tasks, each running in its own thread with a dedicated CUDA stream and vLLM runner, synchronized via concurrency primitives.

\para{Hardware.} We conduct experiments on a server equipped with eight NVIDIA A800-SXM4-80GB GPUs interconnected via 200 GB/s NVLink, two 52-core Intel Xeon Platinum 8470 CPUs, and 1 TB of host memory.

\begin{table}[t]
\centering
\caption{Prompt length across diverse tasks in the evaluation trace.}
\label{tab:length}
\begin{adjustbox}{width=0.5\textwidth}
\begin{tabular}{ccccc}
\toprule
Task & Mean & P99 & Std. & Ratio(\%) \\ 
\midrule
Chatbot (Text)   & 590  & 3040  & 652  & 68 \\
Image Understanding (Image)  & 532  & 2764  & 510  & 8 \\
Web Search (Search) & 5976 & 16635 & 3456 & 20 \\
Summarization (File)   & 6833 & 22390 & 5186 & 4 \\
\bottomrule
\end{tabular}
\end{adjustbox}
\end{table}

\begin{table}[t]
\centering
\caption{Settings of different TTFT SLO requirements.}
\label{tab:slos}
\begin{tabular}{cccccc} 
\toprule
Model & Text & Image & Search & File \\ 
\midrule
Llama3-8B   & 0.25s & 0.5s & 4.0s & 6.0s \\
Qwen2.5-14B & 0.4s & 0.8s & 6.5s & 9.0s \\
Llama3-70B  & 1.0s & 2.0s & 15.0s & 18.0s \\
\bottomrule
\end{tabular}
\end{table}

\begin{figure*}[t]
  \centering
  \includegraphics[width=\textwidth]{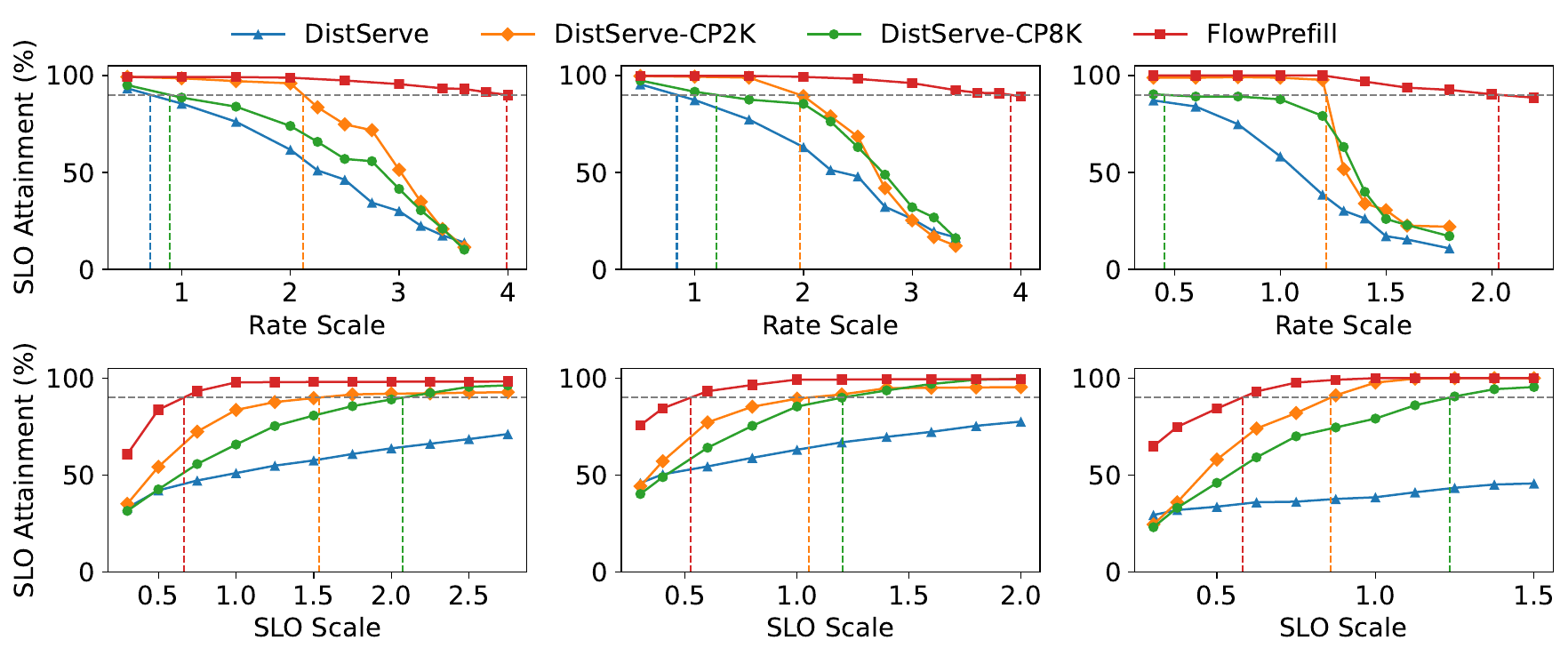}

  \begin{minipage}{0.33\textwidth}\centering
    \text{(a) Llama3-8B}
  \end{minipage}
  \begin{minipage}{0.33\textwidth}\centering
    \text{(b) Qwen2.5-14B}
  \end{minipage}
  \begin{minipage}{0.33\textwidth}\centering
    \text{(c) Llama3-70B}
  \end{minipage}

  \caption{End-to-end performance of three models under varying request rates and SLO requirements on QwenTrace.}
  \label{fig:end2end}
\end{figure*}

\para{Models.} We evaluate \sys{} on three models of varying sizes--Llama3-8B, Qwen2.5-14B~\cite{team2024qwen2}, Llama3-70B~\cite{dubey2024llama3}--using one-, two-, and four-way tensor parallelism respectively. We also use a popular mixture-of-expert model Qwen3-30B-A3B \cite{yang2025qwen3} to show the generalization of our methods.

\para{Workloads.} The evaluation uses a real-world production trace, QwenTrace~\cite{qwen-trace}, which contains four task types with distinct prompt-length distributions (Table~\ref{tab:length}) and timestamped request arrivals (Figure~\ref{fig:intro_distribution}). We preprocess the trace by converting all requests into single-turn queries. To reflect realistic multi-SLO requirements (Table~\ref{tab:slos}), we assign task-specific SLOs based on prior work~\cite{zhong2024distserve, sola} and application semantics. Specifically, text requests have the strictest SLOs for interactive chat, image requests tolerate moderate latency, while search/file requests involve long inputs and thus receive much looser SLOs. As the trace contains only request lengths and does not include the actual prompt contents, we generate random token sequences that conform to the specified lengths for each request.

\para{Metrics.} Since \sys{} performs preemption only during the prefill phase, we focus on TTFT SLO attainment as the primary metric. We define goodput as the maximum sustainable request rate at 90\% SLO attainment. By varying the request rate and latency requirements, we evaluate both the maximum goodput and the minimal SLO the system can meet, which are marked by vertical lines in the curve plots.

\para{Baselines.} We evaluate the effectiveness of \sys{} on both PD-disaggregated and PD-colocated architectures. 1) \emph{Disaggregated Setting}: We compare against the state-of-the-art system DistServe~\cite{zhong2024distserve} under a 1P1D (one prefill, one decode instance) setup. Since DistServe defaults to a First-Come-First-Served (FCFS) policy, we augment its prefill stage with Chunked Prefill (CP)~\cite{chunkedprefill} and an Earliest-Deadline-First (EDF) policy to enable SLO-aware scheduling. We evaluate two variants: DistServe-CP2K (chunk size 2048) and DistServe-CP8K (chunk size 8192). We select 2048 as the primary baseline as it empirically yields the best performance, while 8192 serves as a reference for coarse-grained preemption.
2) \emph{Co-located Setting}: We compare \sys{} against vLLM-0.11.2~\cite{vllm}, which natively supports chunked prefill via temporal multiplexing. This comparison demonstrates the versatility of our method across different serving paradigms.


\subsection{End-to-End Speedup}
We compare the end-to-end performance of \sys{} with the baselines on QwenTrace by varying the request rate and SLO requirements.


\para{SLO attainment vs. request rate.} We evaluate \sys{} under varying request-rate scales for all three models. As shown in the first row of Figure~\ref{fig:end2end}, increasing the request rate raises system load and leads to a gradual decline in SLO attainment, as more requests fail to meet their latency requirements. The vertical lines indicate the maximum goodput. Across all models, \sys{} sustains 4.7$\times$–5.6$\times$ higher request rates than DistServe, and outperforms DistServe-CP2K and DistServe-CP8K by up to 2.0$\times$ and 4.5$\times$, respectively.
This improvement primarily stems from \sys{}'s ability to promptly preempt low-priority requests under heterogeneous SLOs. Upon the arrival of higher-priority requests, event-driven scheduling promptly identifies preemption opportunities, and operator-level preemption suspends ongoing long-input prefill executions with relaxed SLOs to free resources for latency-critical requests, enabling near non-blocking interruption. As a result, latency-sensitive requests are processed without undue delay, which reduces HoL blocking during the prefill phase and prevents SLO violations.

When the request rate continues to increase, all baselines reach their goodput limits. In contrast, \sys{} continues to admit additional requests while satisfying latency constraints. This is enabled by SLO-aware batching, which selects requests that can be batched efficiently without violating their deadlines, and by S-EDF, which proactively deprioritizes requests that cannot meet their deadlines under heavy load. Together, these mechanisms prevent sharp drops in SLO attainment and maintain higher overall performance under high load.

DistServe performs poorly due to its lack of SLO awareness. DistServe-CP2K initially outperforms DistServe-CP8K because its finer-grained preemption better favors latency-sensitive requests. However, as the request rate increases, DistServe-CP2K approaches and may fall behind DistServe-CP8K, since smaller chunks incur higher splitting overhead.

\para{SLO attainment vs. SLO requirements.} The second row of Figure~\ref{fig:end2end} evaluates robustness under different latency requirements. We fix the request rate and linearly scale the latency targets in Table~\ref{tab:slos} using an SLO scale parameter. The vertical lines indicate the minimum SLO targets each system can support. Compared to DistServe-CP2K, \sys{} supports 1.5$\times$--2.3$\times$ tighter SLOs, and compared to DistServe-CP8K, 2.1$\times$--3.1$\times$ tighter SLOs. These gains result from \sys{}’s near non-blocking execution of latency-sensitive requests, which improves SLO attainment and reduces both average and tail latencies. As the SLO scale increases, the SLO attainment of DistServe-CP8K gradually improves and can eventually surpass that of DistServe-CP2K, since larger chunk sizes reduce request fragmentation and tail latency under looser SLO constraints.

\subsection{Ablation Studies}
All ablation studies are conducted using Llama3-8B on a trace segment from QwenTrace.

\begin{figure}[t]
    \centering
    \includegraphics[width=0.49\textwidth]{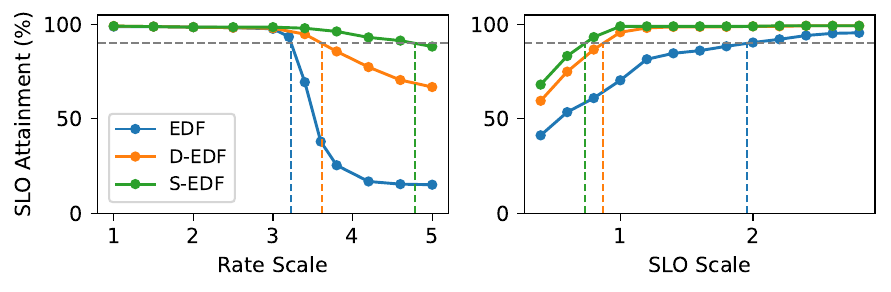}
    \caption{Performance comparison under three scheduling policies.}
    \label{fig:ablation_best_effort}
\end{figure}

\para{Scheduling policy.}
We evaluate the effectiveness of S-EDF by comparing it against two baselines: naive EDF and deadline-aware EDF (D-EDF). For D-EDF, we modify the priority by replacing the numerator with $\sgn(deadline - current\_time)$, such that requests that have already missed their deadlines are assigned the lowest priority. As shown in Figure~\ref{fig:ablation_best_effort}, S-EDF significantly outperforms both baselines in terms of maximum goodput and minimum SLO target. This advantage arises because S-EDF proactively deprioritizes requests that are unlikely to meet their SLOs under high load, thereby preventing a collapse of SLO attainment caused by accumulated queueing delays. In contrast, D-EDF lacks foresight into request feasibility and may waste resources on requests that are destined to violate their SLOs, whereas S-EDF leverages slack estimation to enable more robust scheduling.

\begin{figure}[t]
    \centering
    \includegraphics[width=0.49\textwidth]{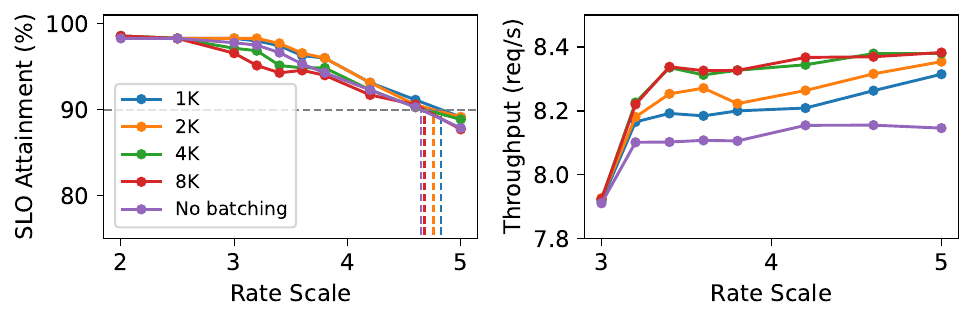}
    \caption{Performance comparison under varying batch token budgets and no batching.}
    \label{fig:ablation_batching}
\end{figure}
\para{SLO-aware batching.}
A batch token budget is imposed to limit batching size. As discussed in \S\ref{sec:takeaway_batching}, exceeding a certain token threshold yields little throughput gain while increasing SLO violation risk. Figure~\ref{fig:ablation_batching} compares SLO attainment and throughput under varying batch token budgets and no batching. The left figure shows that larger budgets increase SLO violation risk but still outperform no batching. The right figure shows that no batching yields the lowest throughput, while larger budgets improve throughput with diminishing returns, with little difference between 4K and 8K. These results suggest that selecting a moderate batch token budget captures the throughput benefits of batching short requests while reducing SLO violation risk, consistent with the analysis above.

\subsection{Runtime Analysis}

\begin{figure}[t]
    \centering
    \includegraphics[width=0.47\textwidth]{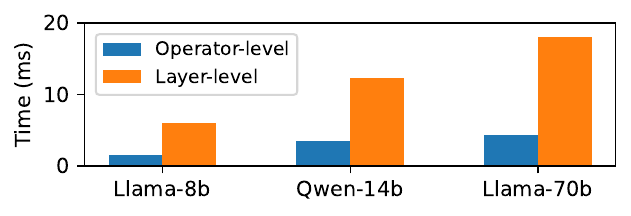}
    \caption{Average preemption blocking time under operator- and layer-level preemption boundaries. Different preemption granularities are realized by placing preemption checks at different execution boundaries.}
    \label{fig:ablation_blocking}
\end{figure}


\para{Preemption blocking time.} 
We define the preemption blocking time as the interval between the moment a higher-priority request sends a preemption signal and the moment the new request can be submitted (i.e., when the ACK is received in Figure \ref{fig:preemption}). 

To quantify the benefits of our approach, we benchmark it against the layer-level preemption boundaries proposed in concurrent work \cite{liao2026laser}. While layer-level scheduling improves granularity, we observed that it remains coarse enough to incur significant delays. As illustrated in Figure~\ref{fig:ablation_blocking}, operator-level preemption reduces the average blocking time by 3.5$\times$--4.2$\times$ compared to layer-level preemption, with all observed latencies remaining below $4.5$ ms. This minimal latency effectively achieves near non-blocking execution for high-priority requests, substantially mitigating HoL blocking during the prefill phase.

While some deployment scenarios may not require such fine-grained preemption, \sys{} allows preemption boundaries to be configured flexibly to accommodate different requirements. Importantly, even when using the finest-grained operator-level preemption, system performance is not adversely affected. Consequently, operator-level preemption is adopted as the default configuration in \sys{}, as it provides robust performance and generality for most production workloads.

\begin{figure}[t]
  \centering
  \includegraphics[width=\columnwidth]{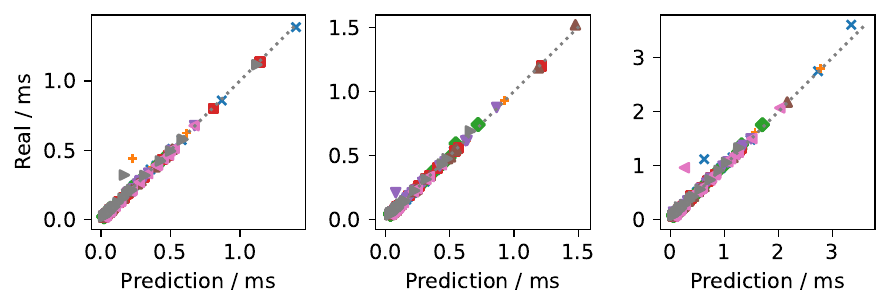}

  \begin{minipage}{0.3\columnwidth}\centering
    \text{(a) Llama3-8B}
  \end{minipage}\hfill
  \begin{minipage}{0.3\columnwidth}\centering
    \text{(b) Qwen2.5-14B}
  \end{minipage}\hfill
  \begin{minipage}{0.3\columnwidth}\centering
    \text{(c) Llama3-70B}
  \end{minipage}
    \caption{Real and predicted TTFT for three models serving a trace segment from QwenTrace at a given request rate.}
  \label{fig:prediction}
\end{figure}

\para{TTFT prediction.} TTFT prediction is required for slack calculator and SLO-aware batching. Our prediction model fits a polynomial to offline prefill profiles, with $x$ as token count and $y$ as predicted TTFT. To validate the model, we record the real TTFT of each request during evaluation. As shown in Figure~\ref{fig:prediction}, most requests have real TTFT close to their predictions, with few outliers, demonstrating accurate prediction even in online scenarios. This accuracy is enabled by the PD-disaggregated setting, where decode execution does not interfere and prefill computation scales nearly linearly with token count, allowing a simple polynomial fit to suffice.

\para{Scheduling cost.}
We count the number of event-driven scheduling decisions during evaluation. Since each request generates at most two events and each event triggers a scheduling round, the total number of scheduling decisions is approximately twice the number of requests and slightly lower when batching occurs. Not every scheduling round results in control commands. Consequently, the number of scheduling rounds that actually trigger submit, preempt, or resume actions is smaller than the total number of scheduling rounds.

\subsection{System compatibility}
All experiments in this section use Llama3-8B, except for the MoE models. Experiments other than the single-SLO scenario use a trace segment from QwenTrace.

\begin{figure}[t]
    \centering
    \includegraphics[width=0.49\textwidth]{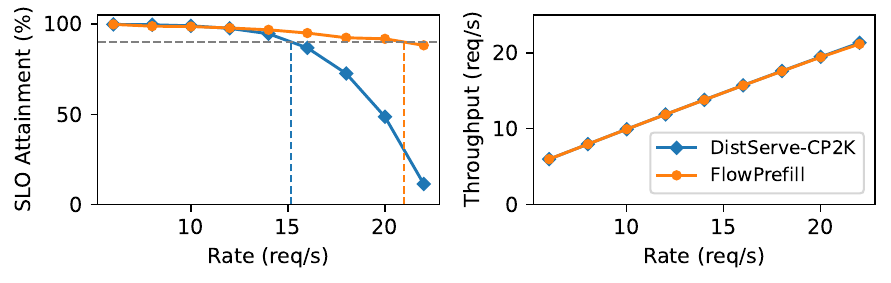}
    \caption{Performance comparison under a single-SLO workload. \sys{} matches baseline throughput, indicating negligible operator-level preemption (§\ref{Operator-Level Preemption}) overhead but delivers better SLO attainment as request rates scale.}

    \label{fig:discussion_sharegpt}
\end{figure}
\para{Single-SLO scenario.} 
We evaluate \sys{} under a single-SLO workload by sampling 500 requests from ShareGPT~\cite{sharegpt} and generating arrivals with a Poisson process at varying rates. All requests share the chatbot SLO in Table~\ref{tab:slos} and have input lengths below 2K, resulting in limited preemption. We therefore compare \sys{} against DistServe-CP2K. 
To evaluate the robustness of \sys{}, we report not only SLO attainment but also throughput, which measures system efficiency regardless of whether SLOs are met. As shown in Figure~\ref{fig:discussion_sharegpt}, \sys{} maintains high SLO attainment while achieving throughput comparable to the baseline, indicating that it preserves responsiveness without sacrificing throughput.

These results demonstrate that the preemption checks used in §\ref{Operator-Level Preemption} do not introduce measurable efficiency loss. Unlike chunk-level or layer-level scheduling, which requires frequent scheduling interventions, operator-level preemption in \sys{} relies on lightweight preemption checks. This design enables fine-grained preemption without interrupting execution, thereby achieving efficient preemption with minimal overhead.

\begin{figure}[t]
    \centering
    \includegraphics[width=0.49\textwidth]{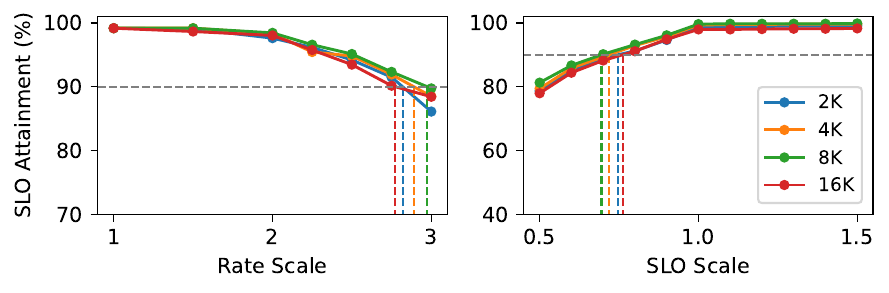}
    \caption{Performance of \sys{} with chunked prefill under different chunk sizes.}
    \label{fig:ablation_mega_chunk}
\end{figure}

\para{Chunked prefill.}
As previously noted, chunked prefill is often employed to reduce HoL blocking in the prefill stage, but it creates a trade-off between responsiveness and throughput that depends on the chosen chunk size. Using large chunks has minimal effect on throughput but is typically impractical because preemption can only occur at coarse boundaries. In contrast, \sys{} supports fine-grained, operator-level preemption, so the blocking time is limited by the maximum runtime of any single operator. For very long inputs, however, one operator can still introduce noticeable latency. In such cases, combining operator-level preemption with chunked prefill reduces individual operator runtimes and further tightens the upper bound on blocking time.

Based on the above observations, we evaluate \sys{} combined with chunked prefill under varying chunk sizes. As shown in Figure~\ref{fig:ablation_mega_chunk}, small chunk sizes (e.g., 2K tokens) provide finer preemption granularity but incur substantial overhead from frequent request splitting, degrading performance. Large chunk sizes (e.g., 16K tokens) reduce splitting overhead but lack sufficient preemption granularity, also hurting performance. An intermediate chunk size balances operator execution time and overall throughput. These results indicate that, when appropriately configured, combining \sys{} with chunked prefill can further improve performance.

\begin{figure}[t]
  \centering
  \includegraphics[width=\columnwidth]{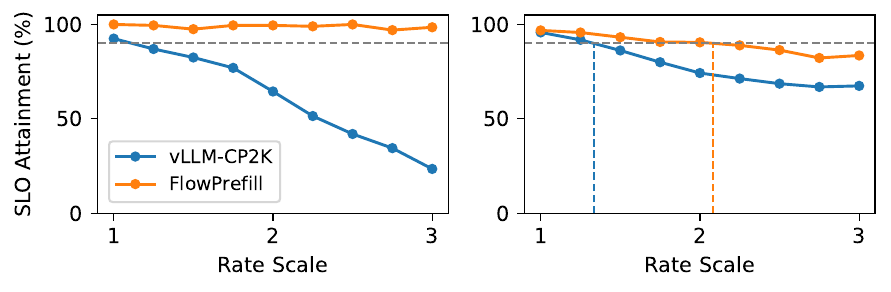}

  \begin{minipage}{0.49\columnwidth}\centering
    \text{(a) TTFT}
  \end{minipage}\hfill
  \begin{minipage}{0.49\columnwidth}\centering
    \text{(b) TBT}
  \end{minipage}

    \caption{Performance comparison under the PD-colocation setting across varying request rates, reporting TTFT and TBT SLO attainment.}
  \label{fig:pd_colocation}
\end{figure}

\para{PD-Colocation.}
Although \sys{} is primarily designed for PD-disaggregated serving, it can also be adapted to a PD-colocation setting. We adopt an intra-PD architecture~\cite{shi2025nexus, lin2025bullet}, where prefill and decode instances are colocated within the same device group. To avoid the complexity of cross-process management, we run on Python~3.14t~\cite{Python3.14t}, which disables the Global Interpreter Lock (GIL) and allows both instances to be managed within a single main process while sharing the same KV-cache logic and memory across devices in the group.

For the PD-colocated evaluation, we benchmark \sys{} against vLLM configured with chunked prefill and an empirically tuned chunk size of 2048 tokens (vLLM-CP2K). Since the colocated setup utilizes only half the number of GPUs compared to the PD-disaggregated configuration, we adjust the SLO constraints accordingly. Specifically, we relax the TTFT SLO to $3\times$ the values in Table~\ref{tab:slos}, and set the Time-Between-Tokens (TBT) SLO to 0.1 seconds for text/image requests and 0.2 seconds for search/file requests.

Figure~\ref{fig:pd_colocation}(a) demonstrates that \sys{} achieves substantially higher TTFT SLO attainment as request rates scale. We also observe positive effects on TBT SLO attainment (Figure~\ref{fig:pd_colocation}(b)). Typically, rising request rates increase prefill demand, causing the interleaving of prefill and decode phases to degrade TBT performance. However, because \sys{}'s adaptive preemption expedites short prefills, it incidentally reduces the duration of interference, improving TBT SLO attainment by up to $1.6\times$.
Although \sys{} primarily targets PD-disaggregated serving, this evaluation demonstrates its robustness under PD-colocation.

\para{Mixture-of-Experts (MoE) models.}
To further validate the generality of our approach, we extend \sys{} to MoE models. In MoE architectures, attention computation uses the same operators as dense models, while the main difference lies in the FFN layer, which includes a gated router and multiple expert FFNs. In practice, this introduces two additional fused operators, \texttt{gate} and \texttt{experts}. Following the \sys{} design, preemption checks are inserted at the boundaries of these operators to enable operator-level preemption. This extension is plug-and-play and can be applied to a variety of model architectures.

\begin{figure}[t]
    \centering
    \includegraphics[width=0.49\textwidth]{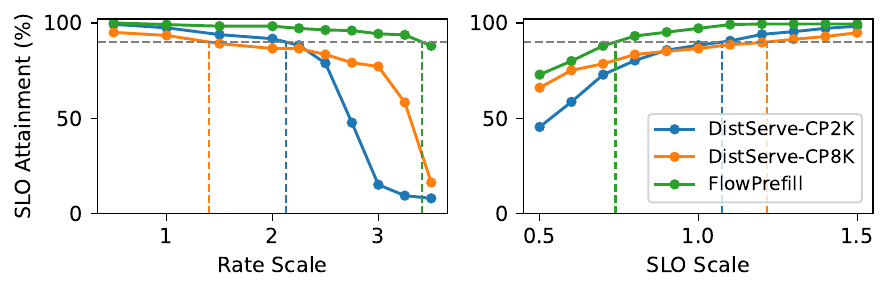}
    \caption{Performance comparison for serving Qwen3-30B-A3B.}
    \label{fig:appendix_moe}
\end{figure}

We evaluate \sys{} on Qwen3-30B-A3B \cite{yang2025qwen3} with 2-way tensor parallelism. Figure~\ref{fig:appendix_moe} compares \sys{} against DistServe-CP2K and DistServe-CP8K. The results are consistent with the end-to-end performance observed for dense models, with \sys{} achieving up to 1.6$\times$ higher goodput and up to 2.4$\times$ tighter SLO attainment. These results demonstrate that \sys{} is compatible with MoE models and remains effective under more complex model structures.

\section{Related Work}
\para{Aggregated LLM Serving Systems.}
Early LLM serving systems~\cite{Faster-Transformer, fang2021turbotransformers, yu2022orca, vllm} optimize performance within a unified execution model. Orca~\cite{yu2022orca} introduces continuous batching to enable per-iteration scheduling, while vLLM~\cite{vllm} improves memory efficiency through fine-grained KV-cache management with PagedAttention. SARATHI~\cite{chunkedprefill} mitigates HoL blocking by interleaving decode with chunked prefill execution. More recent systems~\cite{lin2025bullet, shi2025nexus, hong2025semi-pd} adopt intra-PD architectures that dynamically adjust resources between prefill and decode to improve utilization. While these approaches partially alleviate interference between colocated prefill and decode execution, they struggle to provide consistently high service quality in latency-sensitive online serving scenarios.


\para{Disaggregated LLM Serving Systems.}
Prefill–decode disaggregated architectures~\cite{zhong2024distserve, patel2024splitwise, TetriInfer, qin2024mooncake, epd, feng2025windserve, NVIDIA-Dynamo, liu2025xllm} further improve service quality by decoupling the two phases. DistServe~\cite{zhong2024distserve} jointly optimizes per-phase resource allocation to improve SLO-constrained goodput, while Splitwise~\cite{patel2024splitwise} focuses on cost-efficient placement on heterogeneous resources. TetriInfer~\cite{TetriInfer} stabilizes decode latency under mixed workloads, and Mooncake~\cite{qin2024mooncake} enables efficient disaggregation via a distributed KV-cache store. Although disaggregation removes direct prefill–decode interference, existing systems remain inefficient under heterogeneous SLOs—particularly during the compute-intensive prefill phase, where severe HoL blocking persists. In contrast, \sys{} targets multi-SLO goodput optimization and mitigates prefill-induced HoL blocking through operator-level fine-grained preemption.

\para{LLM Scheduling.}
Existing LLM scheduling policy typically leverages request characteristics, such as input length or SLO targets, to balance latency and throughput. Prior work~\cite{fast-serve, nips-pred, iclr-pred} approximates SJF scheduling using multi-level queues or length prediction. VTC~\cite{vtc} and QLM~\cite{qlm} emphasize fairness and SLO adherence across tenants or models. Other systems~\cite{gong2025past-future, sun2024llumnix, yu2026superinfer} reduce SLO violations via resource-aware dynamic scheduling, and several studies~\cite{sola, slos-serve, slo-aware, lyu2025fairbatching} explicitly address multi-SLO prioritization. Recent approaches~\cite{agrawal2024medha, yu2025prism} use chunked prefill with EDF-style policies to reduce HoL blocking from long requests. However, these methods primarily rely on prioritization and coarse-grained preemption. In contrast, \sys{} directly targets severe HoL blocking caused by long prefill execution and ensures service quality through fine-grained preemption and efficient scheduling.

\section{Conclusion}
This paper presented \sys{}, a goodput-oriented LLM serving system that mitigates prefill-induced HoL blocking under heterogeneous SLO requirements. By decoupling preemption granularity from scheduling frequency through operator-level preemption and event-driven scheduling, \sys{} achieves near non-blocking responsiveness without sacrificing execution efficiency.
Extensive evaluations on real-world production traces demonstrate that \sys{} substantially sustains 4.7$\times$–5.6$\times$ higher request rates than DistServe at the same SLO guarantee. Compared with vLLM, our system also achieves substantial latency improvement. These results indicate that \sys{} provides an effective and practical foundation for efficient multi-SLO LLM serving.


\bibliographystyle{ACM-Reference-Format}
\bibliography{reference}

\appendix


\end{document}